\documentclass[prb,aps,twocolumn]{revtex4}
\usepackage{amsmath,amssymb,graphicx,subfigure,appendix}

\begin{document}

\title{The Prediction of a Gapless Topological ``Haldane Liquid" Phase in a One-Dimensional Cold Polar Molecular Lattice}
\author{J.~P.~Kestner, Bin Wang, Jay D.~Sau, and S.~Das Sarma}
\affiliation{Condensed Matter Theory Center, Department of Physics, University of Maryland, College Park, MD 20742}

\begin{abstract}
We show that ultracold two-component fermionic dipolar gases in an optical lattice with strong two-body on-site loss can be used to realize a tunable effective spin-one model.  Fermion number conservation provides an unusual constraint that $\sum_i \left(S^z_i\right)^2$ is conserved, leading to a novel topological liquid phase in one dimension which can be thought of as the gapless analog of the Haldane gapped phase of a spin-one Heisenberg chain.  The properties of this phase are calculated numerically via the infinite time-evolving block decimation method and analytically via a mapping to a one-mode Luttinger liquid with hidden spin information.
\end{abstract}
\maketitle

\section{Introduction}
We predict theoretically (and support it by numerical calculations) the existence of a completely new quantum topological gapless liquid phase in certain classes of interacting one dimensional systems, which, we believe, can be realized in ultracold fermionic polar molecular gases recently realized in laboratories \citep{Ni08}.  Our predicted novel topological phase can in some sense be construed to be the gapless analog of the well-known Haldane gapped phase that exists in antiferromagnetic spin-one Heisenberg chains \citep{Haldane83}.

Duan \textit{et al.}~\cite{Duan03} have recently shown how to realize a tunable effective spin-half model with ultracold two-component atoms in an optical lattice via superexchange, thus allowing, in principle, the experimental study of quantum magnetism in the controllable atomic systems, but the low temperature scale set by the superexchange is currently a barrier to this program.  Very recent experimental achievement of ultracold gases of dipolar molecules \citep{Ni08} has opened up additional exciting possibilities for realizing designer Hamiltonians due to the extended range of the induced electric dipole-dipole interaction which decays as $1/r^3$.  Efforts are currently underway to load these ultracold molecules into an optical lattice.  In that case one could realize spin models with strong direct off-site interactions, without recourse to the inherently weak superexchange mechanism.  Our proposed new gapless topological phase should be realizable in such a dipolar molecular fermionic optical lattice.

One interesting proposal is to use bosonic polar molecules to realize an incompressible phase with a hidden order similar to the Haldane gapped phase of a spin-one chain \citep{Torre06}.  A recent proposal for explicitly constrained bosons finds similar results \citep{Dalmonte10}.  This intriguing prospect has not yet been experimentally observed, though, as it is specific to bosonic molecules while so far only fermionic dipolar molecules have successfully been realized near degeneracy \citep{Ni08}.  Furthermore, strong two-body loss rates for reactive molecules as in Ref.~\citep{Ni10} would severely limit the stability of states with multiply occupied sites.

In this work, we present a mapping of two-component dipolar fermions to an effective spin-one model.  Rather than being an obstacle, the on-site loss rate is actually a necessary prerequisite for our scheme.  We perform numerical calculations for a one-dimensional (1D) lattice and observe a phase exhibiting hidden order and topological degeneracy. By topological degeneracy, we mean a dependence of the ground state degeneracy on the topology such as occurs in the Haldane gapped phase, where the ground state is four-fold degenerate for open boundary conditions and unique for periodic boundary conditions \citep{Kennedy92}.  In contrast to the ``Haldane insulator" of the bosonic proposal \citep{Torre06}, the phase found here is gapless and compressible, and we refer to it as a ``Haldane liquid."

The layout of the paper is as follows.  In Sec.~\ref{sec:spin1}, we write the effective spin-one Hamiltonian.  In Sec.~\ref{sec:int} we show how the spin-dependent interactions can be experimentally tuned.  Ground state properties of the Hamiltonian are presented in Sec.~\ref{sec:prop}, with numerical results in \ref{subsec:num} followed by an analytical treatment in \ref{subsec:an}.  Experimental detection schemes are briefly discussed in Sec.~\ref{sec:exp}, and we summarize in Sec.~\ref{sec:sum}.  The Appendix contains details of the analysis of Sec.~\ref{subsec:an}.

\section{Effective spin-one description}\label{sec:spin1}
The two fermionic components (\textit{i.e.}, the two spin components) can be provided by two hyperfine states of $^{40}$K$^{87}$Rb molecules \citep{Ospelkaus10}.  The strong on-site loss \citep{Ni10} will give rise to effective hardcore repulsion via a continuous quantum Zeno effect which stabilizes the gas in the presence of the lattice \citep{zeno}.  Thus, after loading the molecules into the lattice, the state will quickly decay to a stable configuration with each site either empty or singly occupied.  For simplicity, we consider geometries such that the interactions are isotropic.  This is not an essential requirement.  The low-energy Hamiltonian is then
\begin{multline}\label{eq:H0}
H = P_S \biggl[ -t \sum_{i,\sigma} \left(c_{i,\sigma}^{\dagger} c_{i+1,\sigma} + \text{h.c.} \right) - \sum_{i,\sigma} \mu_{\sigma} c_{i,\sigma}^{\dagger} c_{i,\sigma}
\\
+ \frac{1}{2} \sum_{i \neq j,\sigma,\sigma'} \frac{V_{\sigma \sigma'}}{|i-j|^3} c_{i,\sigma}^{\dagger} c_{i,\sigma} c_{j,\sigma'}^{\dagger} c_{j,\sigma'} \biggr] P_S ,
\end{multline}
where $\sigma = \uparrow, \downarrow$ denotes the pseudospin state and $P_S$ is the projector onto the subspace with at most one particle per site, \textit{i.e.}, $\sum_{\sigma} c_{i,\sigma}^{\dagger} c_{i,\sigma} \leq 1$.  Note that in 1D the hardcore constraint in conjunction with number conservation breaks the Hilbert space up into disconnected sectors since a given sequence of $\uparrow$ and $\downarrow$ fermions cannot evolve to a different sequence, although the locations of the empty sites may change.

Since the on-site Hilbert space is spanned by three states ($\uparrow$, $\downarrow$, empty), we pursue a mapping to a spin-one description, as in Refs.~\citep{Batista01,Anfossi05}.  To construct spin-one operators which commute properly from the physical fermion operators we introduce a slave fermion representing the empty state, $c_{i,0}$.  Then one can verify that
\begin{equation}
S_{i}^z = c_{i,\uparrow}^{\dagger} c_{i,\uparrow} - c_{i,\downarrow}^{\dagger} c_{i,\downarrow}, \,
S_{i}^+ = c_{i,\uparrow}^{\dagger} c_{i,0} + c_{i,0}^{\dagger} c_{i,\downarrow}, \, S_i^- = \left(S_i^+\right)^{\dagger}
\end{equation}
are effective spin-one operators.

In terms of the slave fermion, the hardcore constraint becomes $\sum_{\eta} c_{i,\eta}^{\dagger} c_{i,\eta} = 1$, where $\eta=\uparrow,\downarrow,0$, and the hopping term in Eq.~\eqref{eq:H0} changes as $-t c_{i,\sigma}^{\dagger} c_{i+1,\sigma} \longrightarrow -t c_{i,\sigma}^{\dagger} c_{i+1,0}^{\dagger} c_{i+1,\sigma} c_{i,0}$.  The constraint can be automatically satisfied by rewriting the Hamiltonian in the form of a spin lattice:
\begin{widetext}
\begin{multline}\label{eq:H1}
H = - t \sum_i \left( S_i^x S_{i+1}^x + S_i^y S_{i+1}^y \right) \left(S_i^z + S_{i+1}^z \right)^2  - \frac{\mu_{\uparrow} - \mu_{\downarrow}}{2} \sum_i S_i^z - \frac{\mu_{\uparrow} + \mu_{\downarrow}}{2} \sum_i \left(S_i^z\right)^2
\\
+ \frac{V_{\uparrow \uparrow} + V_{\downarrow \downarrow} - 2 V_{\uparrow \downarrow}}{8} \sum_{i\neq j} \frac{S_i^z S_{j}^z}{|i-j|^3} + \frac{V_{\uparrow \uparrow} + V_{\downarrow \downarrow} + 2 V_{\uparrow \downarrow}}{8} \sum_{i\neq j} \frac{\left(S_i^z S_{j}^z \right)^2}{|i-j|^3} + \frac{V_{\uparrow \uparrow} - V_{\downarrow \downarrow} }{8} \sum_{i\neq j} \frac{ \left(S_i^z\right)^2 S_{j}^z + S_i^z \left(S_{j}^z\right)^2}{|i-j|^3}
\end{multline}
\end{widetext}
where the $S^z$-dependence of the hopping ensures conservation of $\sum_i \left(S^z_i\right)^2$ (\textit{i.e.}, prevents fermion pair creation and annihilation).

\section{Engineering spin-dependent interactions}\label{sec:int}
\begin{figure}
\subfigure[] {
  \includegraphics[width=.3\columnwidth,height=.3\columnwidth]{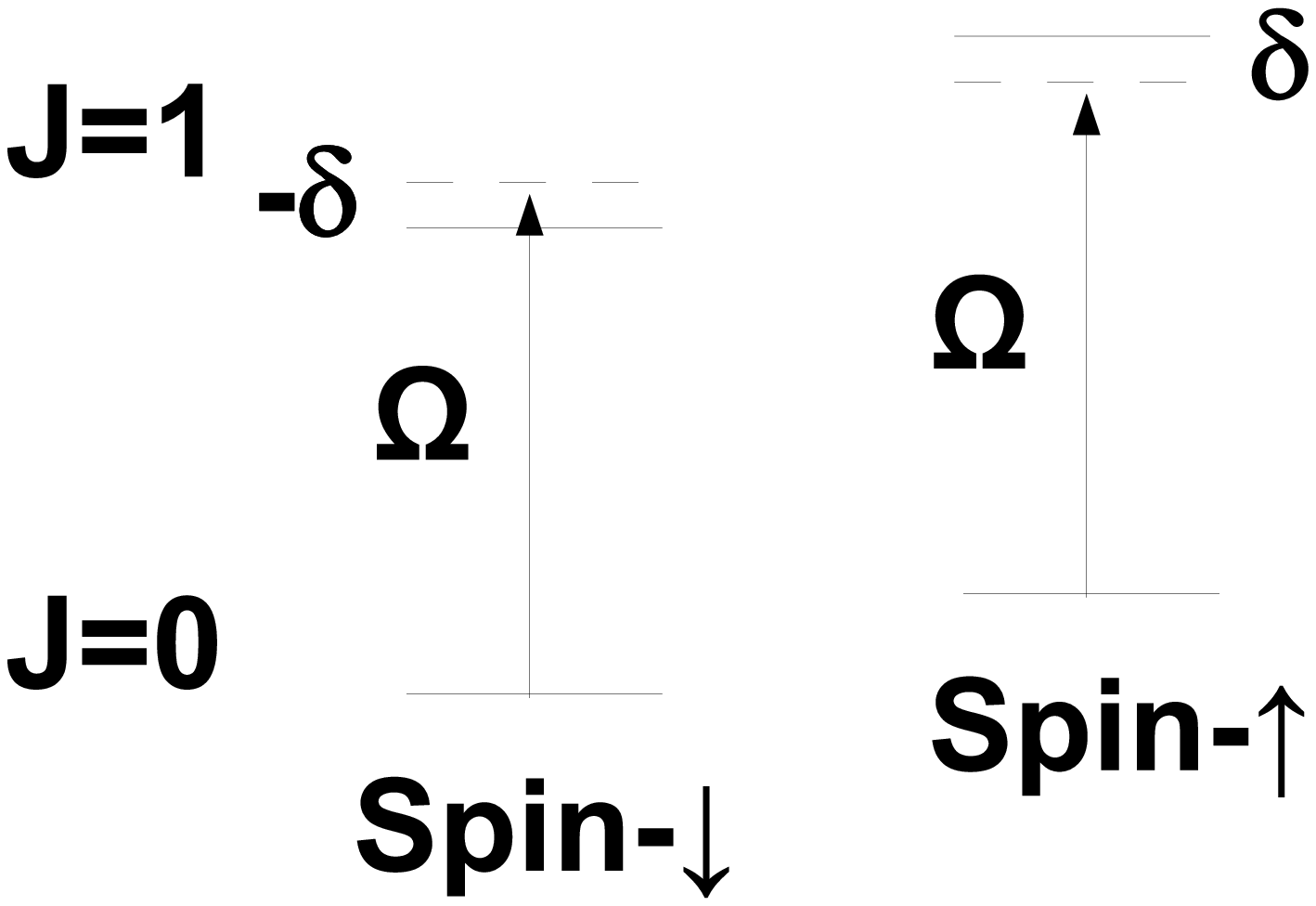}\label{fig:onefield}}
\subfigure[]{
  \includegraphics[width=.6\columnwidth]{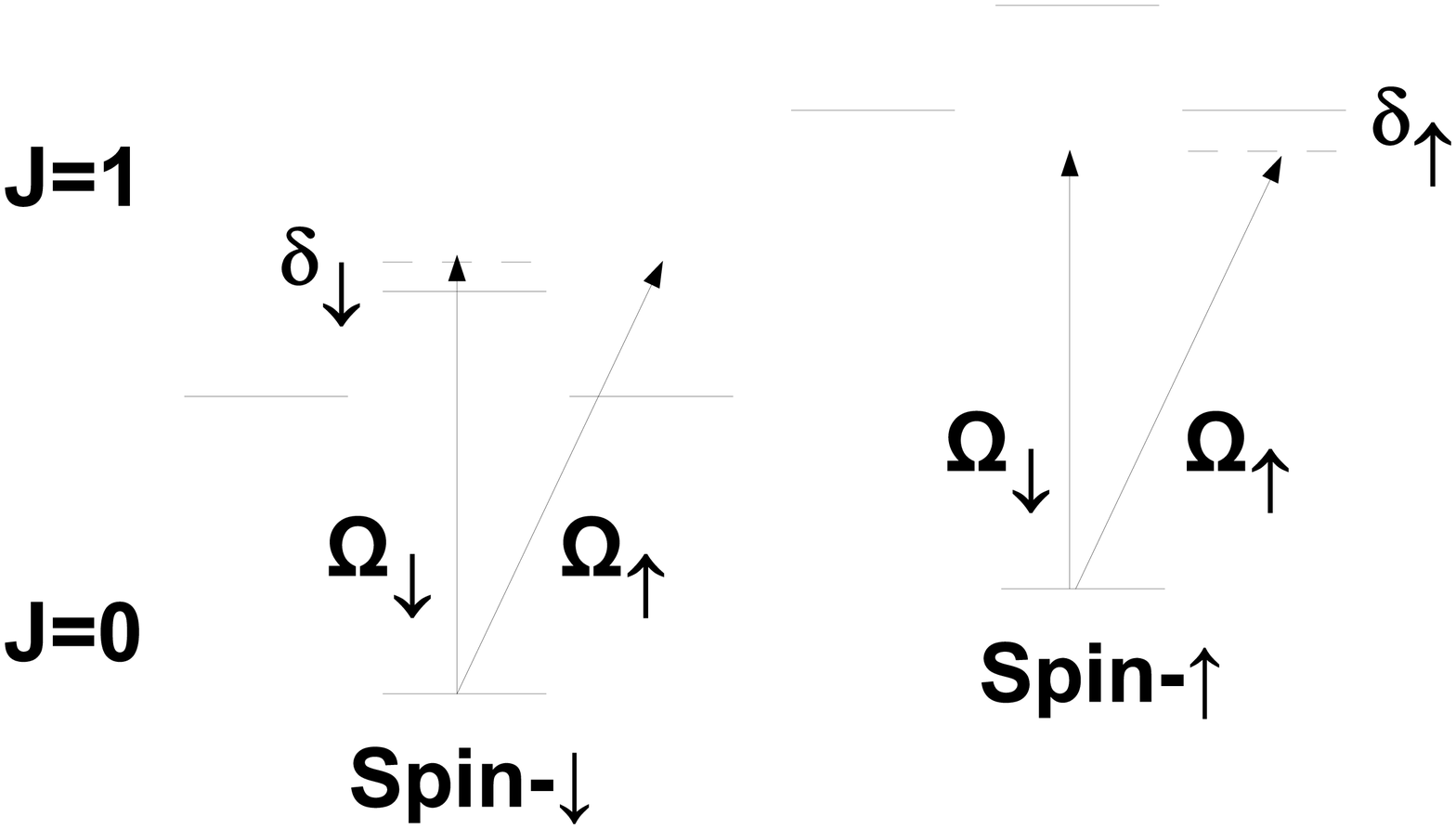}\label{fig:twofields}}
  \caption{Sketch of applied AC fields yielding different interspecies and intraspecies interactions for two hyperfine states in a magnetic field. (a) Scheme yielding $V_{\uparrow \uparrow} = V_{\downarrow \downarrow} = -V_{\uparrow \downarrow}$, (b) Scheme allowing more general control.}\label{fig:fields}
\end{figure}

We now show how the interactions can be tuned.  In the following we keep only nearest neighbor interactions and take the lattice to be one-dimensional, but this can easily be generalized to further interactions and two dimensions.  The Hamiltonian \eqref{eq:H1} becomes especially simple for $V_{\uparrow \uparrow} = V_{\downarrow \downarrow} = -V_{\uparrow \downarrow} = V$, when it reduces to the constrained $t-J_z$ model of Ref. \citep{Batista00} for a stripe segment in a high temperature superconductor.  For $^{40}$K$^{87}$Rb molecules in a typical uniform external magnetic field the resonant microwave transition frequencies between the ground and first excited rotational manifolds are different for different hyperfine species \citep{Aldegunde09}.  (We will consider the Zeeman shift of the noninteracting energy levels to be absorbed into the chemical potentials.)  Thus one could apply an AC field at the average excitation frequency as shown in Fig.~\ref{fig:onefield}, so that the detuning is equal and opposite for the two hyperfine species. If we begin with all molecules in the rotational ground state and the field is turned on adiabatically, in the limit of large intermolecular separation, a molecule will reside in the dressed state adiabatically connected to its ground state.  The difference in the sign of the detuning for the two hyperfine species connects their ground states to different dressed states,
\begin{equation}\label{eq:dressed1}
|\tilde{\sigma} \rangle = e^{i \omega t/2} \cos\theta |0,\sigma \rangle
- s_{\sigma} e^{-i \omega t/2} \sin \theta |1 , \sigma \rangle,
\end{equation}
where $\tan 2\theta = \Omega/\delta$, $\Omega$ is the Rabi frequency, $\delta$ is the detuning, $\omega$ is the field frequency, $|J,\sigma \rangle$ is the bare state in the rotational manifold $J$ and the hyperfine state $\sigma$, and $s_{\sigma}=+1\left(-1\right)$ for $\sigma = \uparrow \left(\downarrow\right)$.  We suppress the $m_J$ label since the field will dominantly couple the $J=0$ state to a single $J=1$ state whose value of $m_J$ is fixed by the polarization.

The resulting effective dipole moments are
\begin{equation}
\mathbf{d}_{\sigma} = \langle \tilde{\sigma} |\mathbf{d}| \tilde{\sigma} \rangle =  -s_{\sigma} \hat{\mathbf{E}} d_{\text{eff}} \cos \omega t,
\end{equation}
where $\hat{\mathbf{E}}$ is the polarization unit vector, $d_{\text{eff}} = d \sin\theta \cos\theta$, and $d$ is the transition dipole moment of the molecule between the ground and first excited state.  The time-averaged intermolecular interaction induced by a field linearly polarized at an angle $\theta_E$ to the $z$-axis is then given by
\begin{multline}
V^{\text{eff}}_{\sigma \sigma'} = \frac{1}{\lambda^3} \langle \mathbf{d}_{\sigma}\cdot \mathbf{d}_{\sigma'} - 3 \left(\mathbf{d}_{\sigma}\cdot \hat{\mathbf{z}} \right) \left(\mathbf{d}_{\sigma'}\cdot \hat{\mathbf{z}} \right) \rangle_t
\\
= \frac{d_{\text{eff}}^2 }{2 \lambda^3} \left(1-3 \cos^2 \theta_E \right) s_{\sigma} s_{\sigma'} .
\end{multline}

More generally, one could control $V_{\sigma \sigma'}$ by carefully tuning a weak DC electric field and the external magnetic field such that the $\Delta m_J=0$ transition for one hyperfine species is nearly resonant with the $\Delta m_J = 1$ transition for the other species, and there are no other transitions near resonance, as shown in Fig.~\ref{fig:twofields}.  Then, applying two microwave AC fields with the same frequency tuned near resonance with the chosen transition, one circularly polarized and one linearly polarized, each species is affected by only one of the fields and the interactions can be tuned individually.  We take the circularly polarized field to propagate at an angle $\theta_{\uparrow}$ relative to the $z$-axis (which is fixed by the direction of the lattice) and the linearly polarized field to have its polarization vector at a polar angle $\theta_{\downarrow}$ relative to the $z$-axis and azimuthal angle $\phi$ relative to the plane formed by the $z$-axis and the propagation vector of the circularly polarized field.  As before, in the limit of large separation, the molecules will reside in the dressed states, but now with $\Omega$ replaced by $\Omega_{\sigma}$, $\delta$ by $\delta_{\sigma}\equiv E_{J=1,\sigma}-E_{J=0,\sigma}-\omega$, and $s_{\sigma}$ by $\text{sgn}\left(\delta_{\sigma}\right)$.  The effective dipole moments, $\mathbf{d}_{\sigma} = \langle \tilde{\sigma} |\mathbf{d}| \tilde{\sigma} \rangle$, are
\begin{equation}
\mathbf{d}_{\downarrow} =  d_{\text{eff},\downarrow} \left(\hat{\mathbf{x}} \sin \theta_{\downarrow} \cos \phi + \hat{\mathbf{y}} \sin \theta_{\downarrow} \sin \phi + \hat{\mathbf{z}} \cos \theta_{\downarrow}  \right) \cos \omega t,
\end{equation}
\begin{equation}
\mathbf{d}_{\uparrow} = d_{\text{eff},\uparrow} \left(\hat{\mathbf{x}} \cos \theta_{\uparrow} \cos \omega t + \hat{\mathbf{y}} \sin \omega t - \hat{\mathbf{z}} \sin \theta_{\uparrow} \cos \omega t \right),
\end{equation}
where $d_{\text{eff},\sigma} \equiv -\text{sgn}\left(\delta_{\sigma}\right) d \sqrt{1-\frac{\delta_{\sigma}^2}{\Omega_{\sigma}^2+\delta_{\sigma}^2}}$.  The time-averaged interaction of molecules on the $z$-axis is then
\begin{align}
V^{\text{eff}}_{\uparrow \downarrow} &= \frac{d_{\text{eff},\uparrow} d_{\text{eff},\downarrow}}{2\lambda^3} \left(2 \sin  \theta_{\uparrow} \cos \theta_{\downarrow} + \cos \theta_{\uparrow} \sin \theta_{\downarrow} \cos \phi \right)
\\
V^{\text{eff}}_{\sigma \sigma} &= s_{\sigma} \frac{d_{\text{eff},\sigma}^2}{2\lambda^3} \left(3 \cos^2  \theta_{\sigma} -1\right).
\end{align}

For intermolecular distances less than or on the order of $r_{\delta} \equiv \left(d^2/\hbar|\delta|\right)^{1/3}$, a full coupled-channel Born-Oppenheimer calculation becomes necessary, as in Refs.~\citep{int}.  However, typically the optical lattice wavelength $\lambda \sim 1\mu$m, which is greater than $r_{\delta}$ for detunings as small as a few kHz, so we do not need to consider the short-range structure of the potential.

\section{Ground state properties}\label{sec:prop}
In the remainder of this work we will consider the ground state of the system, taking the experimental parameters such that $\mu_{\uparrow} = \mu_{\downarrow} = \mu$, $V_{\uparrow \uparrow} = V_{\downarrow \downarrow}=V$ and $V_{\uparrow \downarrow} = - V \cos \chi$.  (We do not discuss the special case $\cos \chi = -1$, corresponding to the trivial situation of spin-independent interactions.)

\subsection{Numerical results}\label{subsec:num}
\begin{figure}[]
  \subfigure[] {
  \includegraphics[height=.45\columnwidth]{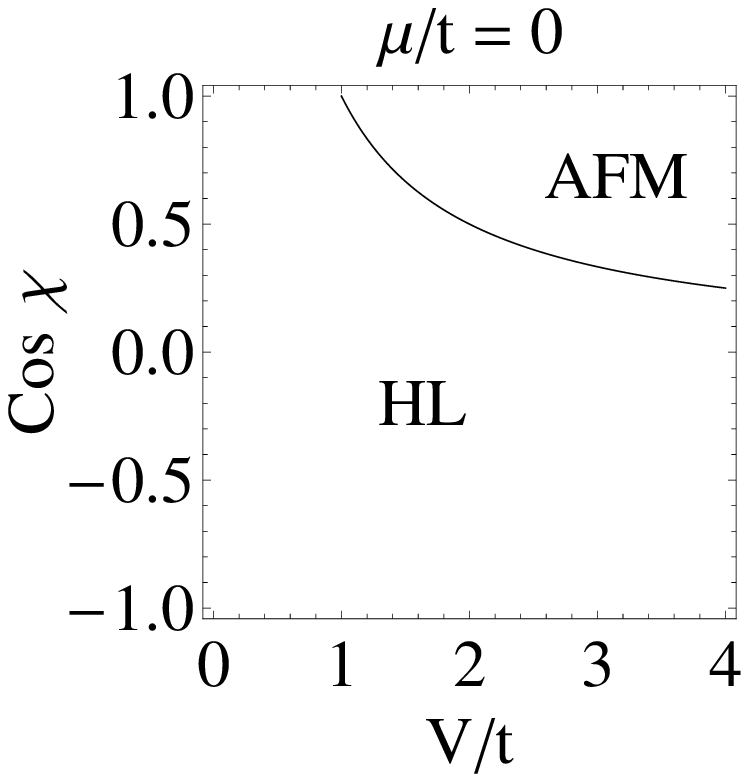}\label{fig:phasediag4mu0}}
  \subfigure[] {
  \includegraphics[height=.45\columnwidth]{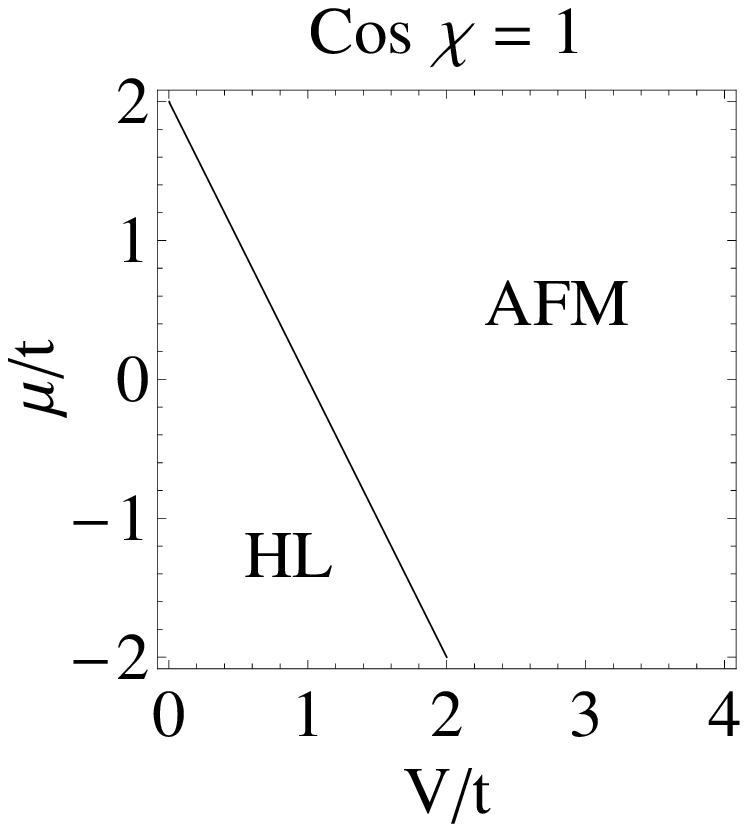}\label{fig:phasediag4chi0}}
\caption{Phase diagrams of Hamiltonian \eqref{eq:H1} in 1D with $\mu_{\uparrow} = \mu_{\downarrow} = \mu$, $V_{\uparrow \uparrow} = V_{\downarrow \downarrow}=V$ and $V_{\uparrow \downarrow} = - V \cos \chi$ and keeping only nearest neighbor interaction.  $V/t$ strictly equal to zero is not included.  The range of chemical potential shown corresponds to filling factors between 0 and 1.}\label{fig:phasediag}
\end{figure}

\begin{figure}[]
  \includegraphics[width=\columnwidth]{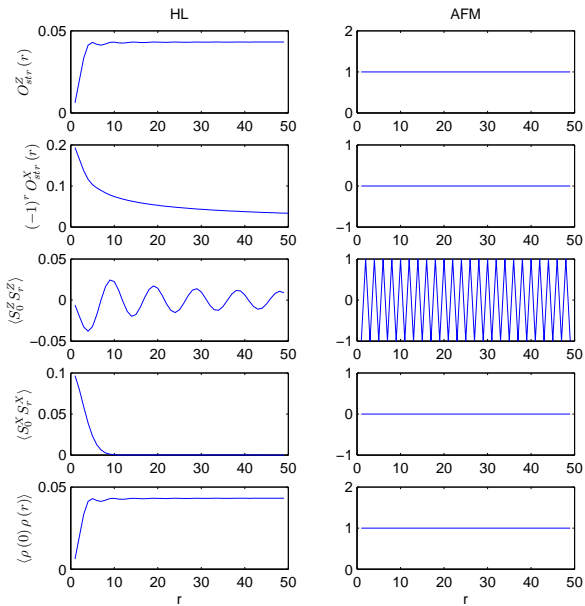}
\caption{Correlation functions calculated numerically with $\cos \chi = 1$ for $V/t=0.1, \mu/t=-1.6$ (HL), and $V/t=1.5, \mu/t=-0.8$ (AFM).}\label{fig:corr}
\end{figure}

We have calculated the ground state properties of the original Hamiltonian \eqref{eq:H0} numerically using the infinite time-evolving block decimation (iTEBD) method \citep{Vidal07}.  The phases shown in Fig.~\ref{fig:phasediag} are defined by the structure of the entanglement spectrum, as we discuss below, and characterized by the correlation functions shown in Fig.~\ref{fig:corr}.

For $V>0$ the ground state clearly must lie in the sector of the Hilbert space containing states like $\uparrow$0$\downarrow$$\uparrow$00$\downarrow\cdot\cdot\cdot$, \textit{i.e.}, where the hyperfine state of each molecule is different from that of the previous molecule, regardless of how many empty sites lie in between.  This is long range order in the string correlator \citep{denNijs89},
\begin{equation}
O_{\text{str}}^{\alpha} \left(|i-j|\right) \equiv \langle -S_i^{\alpha} \exp \left(i\pi \sum_{l=i+1}^{j-1} S_l^{\alpha} \right) S_j^{\alpha} \rangle,
\end{equation}
with $\alpha = z$.  (We do not consider the special case of $V=0$, which is separated from the $V>0$ phase by the closing of the energy gap between the low-lying $z$-string ordered states and states which break this order.)  When $V\cos\chi > t-\mu/2$ \citep{note}, the entanglement spectrum is nondegenerate and every lattice site is occupied, so the ground state is an antiferromagnetic insulator (AFM), and must be doubly degenerate.  In a system with a fixed number of molecules less than the number of lattice sites, this will be a phase separated region.

When $V\cos\chi < t-\mu/2$, there are vacancies and the entire entanglement spectrum is exactly doubly degenerate, which is an indication of a nontrivial topological state \citep{Pollman10}.  However, in contrast to the ``Haldane insulator" of Ref.~\citep{Torre06}, here the suppression of pair creation/annihilation processes has removed the gap to the longitudinal magnon mode, as can be surmised from the power law decay of the spin-spin correlations along $z$ shown in Fig.~\ref{fig:corr}, while the transverse modes remain gapped \citep{Schulz86}.  Furthermore, the string correlator along $x$ no longer displays true long range order, but decays with a power law.  Algebraic decay of string correlation has previously been found in the Hubbard model with infinite onsite repulsion (i.e., Luttinger liquid with Heisenberg exchange) \citep{Kruis04}.  In the present case, though, both long range and quasi-long range string order exist simultaneously.  Most importantly, we shall show below that the doubly degenerate entanglement spectrum is the direct result of topological degeneracy.  We refer to this phase as a ``Haldane liquid" (HL).

\subsection{Analytical results}\label{subsec:an}
The mapping from hard-core spin-half fermions to a spin-one model provides a suggestive and intuitive language to describe the system, but it is difficult to explain the behavior of the HL phase in this language.  However, given an underlying string-type spin order, one can alternatively make a mapping from the original hardcore spin-half fermions to a spinless Luttinger liquid with a prescription for how to access the spin information from the density, similar to the approach in Ref.~\citep{Batista00,Kruis04}.  This provides analytical solutions for the low-energy physics.  We heuristically present the mapping below.  A more detailed presentation is given in the Appendix.

\subsubsection{Effective spinless Luttinger liquid description}
Consider a lattice with sites $0,1,...,N-1$ and periodic boundary conditions loaded with $n\leq N$ fermions at sites $\mathbf{r} = \{r_1 < r_2 < ... < r_n\}$ bearing spins $\mathbf{\sigma} = \{\sigma_1,\sigma_2,...\sigma_n\}$.  Restricting ourselves to the physically relevant string ordered subspace, eliminating the spin information leaves a spinless attractive fermion system.  Any eigenstate of the spinful Hamiltonian, $H=T+V$, in this subspace can be written as a superposition
$
\Psi\left(\mathbf{r},\mathbf{\sigma}\right) = \sum_{\gamma = \uparrow,\downarrow}\alpha_{\gamma} \psi_{\gamma} \left(\mathbf{r},\mathbf{\sigma}\right) = \sum_{\gamma}\alpha_{\gamma} \xi_{\gamma}\left(\mathbf{\sigma}\right) \phi \left(\mathbf{r}\right)$,
where the two terms correspond to the two possible realizations of string order, \textit{i.e.}, choice of the first fermion's spin.  This choice uniquely determines all the spins, so the state separates into spin and density parts, and $\phi \left(\mathbf{r}\right)$ is an eigenstate of the spinless attractive fermion Hamiltonian, $\tilde{H} = \tilde{T} + V$.  (Note that the interaction in the string ordered subspace is independent of the two spin states.)

If the lattice is packed, the two string ordered sectors are disconnected and all eigenstates are doubly degenerate.  For $n < N$, though, the hopping term, $T$, connects states $\psi_{\uparrow} \left(\mathbf{r},\mathbf{\sigma}\right)$ and $\psi_{\downarrow} \left(\mathbf{r},\mathbf{\sigma}\right)$ by taking a particle from site $N-1$ to site $0$ (or vice versa) via the periodic (or more generally, phase twisted) boundary conditions.  In this case the symmetry of the Hamiltonian requires the eigenstates to have definite parity, $p=\pm 1$, under ``time reversal" (TR), \textit{i.e.}, $\uparrow \Leftrightarrow \downarrow$, so the eigenstates are unique.

\begin{figure}[]
  \includegraphics[width=\columnwidth]{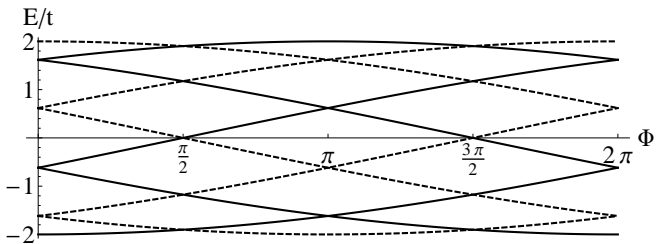}
\caption{Energy vs. flux for a 5-site ring with 4 string ordered fermions.  Solid (dashed) lines are TR even (odd) states.}\label{fig:fivesite}
\end{figure}

Consistency of the spinful and spinless Schrodinger equations requires $T \Psi\left(\mathbf{r},\mathbf{\sigma}\right) = \sum_{\gamma} \alpha_{\gamma} \xi_{\gamma}\left(\mathbf{\sigma}\right) \tilde{T} \phi \left(\mathbf{r}\right)$.  This is clearly satisfied for hops that do not move a particle between sites $N-1$ and $0$.  Consistency for hops between sites $N-1$ and $0$ requires $e^{i\Phi} = p e^{i\tilde{\Phi}}$, where $\Phi$ ($\tilde{\Phi}$) is the boundary phase twist of the spinful (effective spinless) system.  So for TR even (odd) states, we may describe the spinful state in terms of an effective spinless state under boundary conditions with the same (an additional $\pi$) phase twist.  The meaning of this is readily seen in that $N$ hops are sufficient to cycle the density around the ring back into its original configuration, but the resulting spin part of the wavefunction is time-reversed.  Thus, TR odd states acquire a $\pi$ phase after one circuit.  One interesting ramification is that, if an effective flux is introduced, the ground state energy as a function of flux has half the period of a spinless system, as illustrated in Fig.~\ref{fig:fivesite} by exact diagonalization in the string ordered Hilbert space.

Given the above prescription to obtain the string ordered spinful wavefunction from a spinless one, one obtains (see Appendix for details) algebraic decay of the $x$ string, $z$ spin-spin, and density-density (deviation from mean) correlation functions in the HL phase from Luttinger liquid theory:
\begin{align}
&O_{\text{str}}^{x} \left(r\right) \sim \frac{\left(-1\right)^r}{4} \left(\frac{\lambda}{r}\right)^{1/2K},
\\
&\langle S_0^z S_r^z \rangle \sim \left[\frac{k_F^2}{\pi^2}+\frac{k_F}{2\pi^2\lambda}+\frac{1}{\left(2\pi\lambda\right)^2}\right] \cos k_F r \left(\frac{\lambda}{r}\right)^{K/2} \!\!,
\\
&\langle \rho \left(0\right) \rho\left(r\right) \rangle - \langle \rho\left(0\right) \rangle^2 \sim -\frac{K}{2\pi r^2} \, ,
\end{align}
where $K>1$ is the Luttinger constant.  We have checked that the numerical calculations give consistent results for $K$ when comparing the various correlation functions.
Also, the exponential decay of $\langle S_0^x S_r^x \rangle = \sum_{\gamma} \langle c^{\dagger}_{0,\gamma} c_{r,\gamma} \rangle$ is readily understood since one cannot propagate one particle past another without destroying the string order, so the correlation function must exponentially decay with a characteristic length set by the average interparticle spacing.

\subsubsection{Entanglement spectrum degeneracy}
From the above arguments it is clear that under open boundary conditions, the ground state must always be doubly degenerate, as also shown in Ref.~\citep{Batista00}.  However, for periodic boundary conditions it may be unique.  To link the double degeneracy of the entanglement spectrum to this topological degeneracy, consider the reduced density matrix for a block of sites, $A$, obtained by tracing over the remaining subsystem, $B$,
\begin{multline}
\rho_A \left(\mathbf{r_A},\mathbf{\sigma_A}; \mathbf{r_A'}, \mathbf{\sigma_A'}\right) =
\\
\sum_{\mathbf{r_B},\mathbf{\sigma_B}} \Psi\left(\mathbf{r_A},\mathbf{r_B};\mathbf{\sigma_A},\mathbf{\sigma_B}\right) \Psi^{\ast} \left(\mathbf{r_A'},\mathbf{r_B};\mathbf{\sigma_A'},\mathbf{\sigma_B}\right).
\end{multline}
Since a given number of particles and realization of string order in $B$ specifies the number and realization of string order in $A$, the above decomposes as
\begin{equation}
\rho_A \left(\mathbf{r_A},\mathbf{\sigma_A}; \mathbf{r_A'}, \mathbf{\sigma_A'}\right) =  \underset{n_A, \gamma}{\oplus} \rho_A^{\left(n_A, \gamma\right)} \left(\mathbf{r_A}, \mathbf{r_A'}\right).
\end{equation}
In the nondegenerate case, the states have definite TR parity, so $\rho_A^{\left(n_A, \uparrow\right)} = \rho_A^{\left(n_A, \downarrow\right)}$ and the entanglement spectrum is exactly doubly degenerate.

\section{Experimental detection}\label{sec:exp}
To probe the HL phase experimentally, one should initialize the system in the $z$ string ordered sector (\textit{e.g.}, by starting with a 3D lattice and adiabatically turning on a strong cylindrical trap).  Local correlators are directly observable.  For instance, $\langle S_0^x S_r^x \rangle$ can be probed by measuring the physical pseudospin-dependent single-fermion correlators, $\sum_{\sigma} \langle c_{i,\sigma}^{\dagger} c_{j,\sigma} \rangle$, via pseudospin-dependent time-of-flight measurement.  Likewise, $\langle S_0^z S_r^z \rangle$ can be probed by measuring the density-density correlator, $\langle \left(\rho_{\uparrow} \left(0\right)-\rho_{\downarrow} \left(0\right)\right) \left(\rho_{\uparrow} \left(r\right)-\rho_{\downarrow} \left(r\right)\right) \rangle$, via pseudospin-dependent time-of-flight noise correlations or repeated \textit{in situ} imaging with single-site resolution.  Although finite temperature introduces an exponentially decaying envelope to the gapless correlation functions, at sufficiently low temperature the algebraic decay is dominant up to some thermal length.

The entanglement spectrum and string correlators are not as directly accessible.  However, one can observe other consequences of the topological degeneracy.  For example, in principle, when realizing the simplest case of $\chi = 0$ with the arrangement of Fig.~\ref{fig:onefield}, one could use a rotating ring-shaped optical lattice \citep{Amico05}.  The half periodicity of the ground state energy as a function of rotation frequency (see Fig.~\ref{fig:fivesite}) compared to the corresponding system of spinless fermions could then serve as a signature of the phase.

\section{Summary}\label{sec:sum}
We have shown that an optical lattice loaded with two-component dipolar fermions can realize an effective spin-one model.  Our proposal allows for -- in fact, requires -- experimentally realistic conditions of strong two-body loss.  In one dimension, we have provided numerical and analytical calculations to show that this model contains a novel gapless topological phase.

This work is supported by AFOSR-MURI, DARPA-QUEST, and ARO-DARPA-OLE.

\appendix\label{app}
\section{}
\subsection{Hamiltonian and Hilbert space decomposition}
Consider a set of spin-1/2 Ising hard core fermions on a 1D lattice with
 anti-ferromagnetic nearest neighbor interactions.  (This model has previously appeared in a different context \citep{Batista00}.)
 Let us suppose the 1D lattice is a finite size ring
with some number of sites $N$.
The Hamiltonian may be written as
\begin{multline}
H=\sum_{r,\sigma=\pm 1}[c_{r+1,\sigma}^\dagger (1-n_{r+1}) c_{r,\sigma}+h.c-\mu c_{r,\sigma}^\dagger c_{r,\sigma}]
\\
-U\sum_{\alpha,\beta,\gamma=\pm 1}\sigma_{z,\alpha\alpha}\sigma_{z,\beta\beta} c_{r+\gamma,\alpha}^\dagger c_{r,\beta}^\dagger c_{r,\beta} c_{r+\gamma,\alpha}\label{ham1}
\end{multline}
where $n(r)=\sum_\sigma c^\dagger_{r,\sigma}c_{r,\sigma}$.
 Furthermore let us suppose that the chemical potential $\mu$
is tuned such that the system has an even number $(n)$ fermions in the ground state.

Each particle in this system can be labeled by a composite variable $y=(r,\sigma)$ where $0\leq r\leq N$ is the position of the particle and $\sigma=\pm 1$ is the spin state. For our purposes we will need to define an order
 between 2 values $y_1=(r_1,\sigma_1)$ and $y_2=(r_2,\sigma_2)$ based
on the order of $r_1$ and $r_2$ i.e. $y_1<y_2 \equiv r_1<r_2$.
The Hilbert space $\mathcal{H}$ of the system above is spanned by the
 ket states
$\mathcal{H}=[|0\leq y_1<y_2<\dots<y_n\leq (N-1) \rangle]$.
 The hard core
 constraint together with the Ising constraint on the spins
 leads to an important conserved quantity in the Hamiltonian given by the
 sequence $\Sigma=(\{\sigma_j\}_{j=1,\dots,n})(mod R)$ i.e the spin sequence is conserved modulo cyclic permutations $R$.
 Thus the Hilbert space may be decomposed into a
direct sum
\begin{equation}
\mathcal{H}=\oplus_\Sigma \mathcal{H}_\Sigma
\end{equation}
where $\Sigma$ represents various equivalence classes of
sequences of spins of the $n$ atoms.
Each Hilbert space $\mathcal{H}_\Sigma$ for a given spin configuration
has a simple structure $\mathcal{H}_\Sigma=[|0\leq r_1<r_2<\dots<r_{n-1}<r_n\leq N\rangle]$
 such that the spin degree of freedom is now implicit.
The spin-degree of freedom can be restored given the value of $\Sigma$.

\subsection{Low energy wave-functions}
After removing the spin-degree of freedom, each Hilbert space and
 corresponding Hamiltonian have the form of spin-less hard core particles
 moving on a 1D lattice. The interactions between these particles is
 attractive
if the spin configuration is in the equivalence class
$\Sigma=(+,-,+,-,\dots)$
i.e the state is antiferromagnetic. The ground state is therefore
purely antiferromagnetic. There are 2 antiferromagnetic
spin sequences in the equivalence class
 namely, $\Sigma_+=(+,-,+,-,\dots)$ and
 $\Sigma_-=(-,+,-,+,\dots)$.

 Thus a wave-function in $\mathcal{H}_\Sigma$
 can be decomposed into 2 parts $\Psi=\Psi_++\Psi_-$
 such that $\Psi_\alpha\in \mathcal{H}_{\Sigma_\alpha}$.
To understand the condition under which $\Psi$ is an
eigenstate, we apply the Hamiltonian in Eq.~\eqref{ham1} to $\Psi$.
For convenience we consider all position variables lumped together
as $\mathbf{r}=(r_1,\dots,r_n)$. All terms in the Hamiltonians under
 consideration are diagonal (and therefore like a potential energy term)
 other than
the hopping term. The composite vector $\mathbf{r}$ is taken to be in a bounded
region such that $r_i<r_{i+1}$.  The hopping terms $\mathcal{T}$
for most of the points
allow only hopping inside this bounded region. Because of periodic
boundary conditions, the hopping $\mathcal{T}$ also connects the boundary
points with $\mathbf r$ such that $\mathbf r=(0<r_1<\dots<r_n=N-1)$ and
$\mathbf r=(0=r_1<\dots<r_n<N-1)$. We refer to these hyperplanes as
$S_-$ and $S_+$ respectively in parameter space.
 The Schrodinger equation takes the abstract form
\begin{equation}
\mathcal{T} \Psi(\mathbf y)=(E-U(\mathbf y))\Psi(\mathbf y)\label{se1}
\end{equation}
where $\mathbf{y} \equiv 0\leq y_1<y_2<\dots<y_n\leq (N-1)$, $\mathcal{T}$  are the allowed
nearest neighbor hopping operators in the multi-dimensional space
 and $E$ is the energy eigenvalue.

 Since the spin ordering
of each state is alternating, the potential energy is independent of
the spin part i.e. $U(\mathbf y)=U(\mathbf r)$.  Although the kinetic energy can mix the two spin sequences, it is also symmetric, so for $n<N$ the eigenstates are nondegenerate and have definite parity under time-reversal.  Therefore, for $n<N$, given a state $\Psi\in \mathcal{H}_{\Sigma}$ with time-reversal parity $p=\pm 1$, we can define a spin-less
wave-function $\psi(\mathbf r)$ defined by
\begin{multline}
\psi(\mathbf r)=\Psi(0\leq (r_1,+)<(r_2,-)<\dots<(r_n,-) < N )
\\
+p\,\Psi(0\leq (r_1,-)<(r_2,+)<\dots<(r_n,+) < N )
\end{multline}
which satisfies the Schrodinger equation
\begin{equation}
\tilde{\mathcal{T}} \psi(\mathbf r)=(E-U(\mathbf r))\psi(\mathbf r)\label{se2}
\end{equation}
where $\tilde{\mathcal{T}}$ is the spin-less hopping operator on the
lattice.
The above Schrodinger equation arises from the
spin-less version of the original Hamiltonian in Eq.~\eqref{ham1},
\begin{equation}
H_{f}=\sum_{r}[c_{r+1}^\dagger  c_{r}+h.c-\mu c_{r}^\dagger c_{r}]-U\sum_{\gamma=\pm 1} c_{r+\gamma}^\dagger c_{r}^\dagger c_{r} c_{r+\gamma},
\end{equation}
which is the Hamiltonian for spin-less fermions on a lattice. Here
we also choose $\mu$ such that the ground state has $n$ fermions.
Note that the
hard core constraint is automatically enforced. Similar to our original
model the Hilbert space can be described by wave-functions
 $\psi(0\leq r_1<r_2<\dots<r_n\leq N-1)$.

Conversely, the spin-ful wave-function can be obtained given
the spinless one $\psi$ by
\begin{equation}\label{eq:psi}
\Psi(\mathbf y)=\sum_{\alpha=\pm 1}\zeta_{\alpha}
\chi_\alpha(\mathbf \sigma)\psi(\mathbf r)
\end{equation}
where $\zeta_{+}=p\,\zeta_{-}$ and $\chi_\alpha$ is the spin part of the wave-function
and is given by
\begin{equation}
\chi_\alpha (\mathbf \sigma )=\prod_{j=1,\dots,n}\delta(\sigma_j+\alpha  (-1)^j).
\end{equation}
The separable form of the wave-function is also obtained for the large-$U$ limit of the Hubbard model by Refs.~\citep{Woynarovich82,Ogata90} and used by Ref.~\citep{Kruis04} in a similar program.  Inserting Eq.~\eqref{eq:psi} into
the right-hand side of Eq.~\eqref{se1}, we obtain
\begin{equation}
\mathcal{T} \Psi(\mathbf y)=\sum_{\alpha=\pm 1}\zeta_{\alpha}
\chi_\alpha(\mathbf \sigma)\tilde{\mathcal{T}}\psi(\mathbf r).\label{se3}
\end{equation}

The validity of Eq.~\eqref{se3} is easy to see in the interior
of the space $\mathbf r$, since the hopping term does not transfer
any particles across the boundary at $r=0$.
For points $\mathbf y$ containing particles at the boundaries $r=0$ or $r=N-1$,
the equation becomes
\begin{equation}
\sum_{\alpha=\pm 1}\zeta_{-\alpha}
\chi_\alpha(\mathbf \sigma)e^{i\theta}\psi'(\mathbf r)=\sum_{\alpha=\pm 1}\zeta_{\alpha}
\chi_\alpha(\mathbf \sigma) e^{i\tilde{\theta}}\psi'(\mathbf r),
\end{equation}
where $\theta$ ($\tilde{\theta}$) is the phase twist at the boundary of the spin-ful (spin-less) lattice and $\psi'(\mathbf r)$ denotes the density configuration after the hopping operator is applied.  So we see that for $\Psi$ time-reversal even, we may describe the system in terms of a spin-less wave-function by using Eqs.~\eqref{eq:psi} and \eqref{se2}, where the phase twist on the boundary of the spin-less Hilbert space is the same as that of the spin-ful Hilbert space.  But for $\Psi$ time-reversal odd, the boundary conditions used in Eq.~\eqref{se2} to obtain the spin-less wave-function must have an additional $\pi$ phase twist.

The above arguments are specifically for $n<N$.  For the filled lattice $n=N$, the
hopping term does not act on the wave-function, and the eigenstates are two-fold degenerate.

\subsection{String order}
The Hilbert space $\mathcal{H_A}$
of the low-energy anti-ferromagnetic states
can be compactly characterized by defining a spin-1 operator $\mathbf S_r$
at each site $r$ of the lattice such that
 $S_{r_j,z}=\sigma_j$ for filled sites $r_j$ and $S_{r,z}=0$ for all
 other sites. With this definition, the Hilbert space $\mathcal{H_A}$
is characterized by the expectation value of the
 string order of $S_z$ operators
defined as
\begin{equation}
O_{\text{str}}^{z}(r<r')=S_{r,z}\prod_{r<p<r'}e^{i\pi S_{p,z}}S_{r',z}.
\end{equation}
The operator $O_{\text{str}}^{z}(r<r')$ is non-zero only if both
$r$ and $r'$ are occupied. The string term in the middle
 $\prod_{r<p<r'}e^{i\pi S_{p,z}}$ computes the parity of the number of occupied sites in-between $r$ and $r'$. Thus $O_{\text{str}}^{z}(r<r')=n(r)n(r')$
 where $n(r)$ is the occupation operator at site $r$.
 The asymptotic expectation value
\begin{equation}
\langle O_{\text{str}}^{z}(r<r')\rangle=\langle n(r)n(r')\rangle\sim \langle n(r)\rangle^2
\end{equation}
 for large $r'-r$.

\subsection{Entanglement spectrum degeneracy}
The entanglement spectrum of a block $A$ of length $l$ starting
from $r=0$ to $r=l-1$ of the ring. The rest of the ring
we refer to as $B$. In the spin-1 operator notation, the Hilbert space
is specified by the states $|\{S_{r,z}\}_{r=0,\dots,N-1}\rangle$ where
$S_{r,z}$ are the spin-1 operators at position $r$. The ground state
wave-function can be correspondingly written as
$\Psi_0(\{S_{r,z}\}_{r=0,\dots,N-1})$. The reduced density
matrix is defined as
\begin{widetext}
\begin{equation}
\rho_{\mathcal{A}}(\{S_{r,z}\}_{r=0,\dots,l};\{S'_{r,z}\}_{r=0,\dots,l})=\sum_{\{S_{r,z}\}_{r=l+1,\dots,N-1}}\Psi(\{S_{r,z}\}_{r=0,\dots,N-1})\Psi^*(\{S'_{r,z}\}_{r=0,\dots,l}\{S_{r,z}\}_{r=l+1,\dots,N-1}).
\end{equation}
\end{widetext}
The spin variable $S_{r,z}=0$ for empty sites. Therefore the
spin-configurations of non-zero weight must satisfy
\begin{equation}
\sum_r S_{r,z}^2=\sum_{0\leq r\leq l} S^{'2}_{r,z}+\sum_{l< r\leq N-1} S_{r,z}^2=N-n.
\end{equation}
Thus $\sum_{0\leq r\leq l} S_{r,z}^2=\sum_{0\leq r\leq l} S^{'2}_{r,z}=n_A$ and the reduced density matrix is diagonal in the number
of fermion $n_A$ in the block $A$.
Therefore the reduced density matrix can be written in the fermion
notation as
\begin{widetext}
\begin{multline}
\rho_{\mathcal{A}}(0\leq y_1<\dots<y_{n_A}\leq l;0\leq y'_1<\dots<y'_{n_A}\leq l)=\sum_{l<y_{n_A+1}<\dots<y_{n}<N} \Psi(0\leq y_1<\dots<y_{n}<N)
\\
\times\Psi^*(0\leq y'_1<\dots<y'_{n_A}\leq l< y_{n_A+1}<\dots<y_{n}<N).
\end{multline}
\end{widetext}
From the above it follows that $\rho_{\mathcal{A}}(0\leq y_1<\dots<y_{n_A}\leq l;0\leq y'_1<\dots<y'_{n_A}\leq l)\propto \delta_{\sigma_0,\sigma'_0}$.
Furthermore, for the case with fermion vacancies, $\Psi$ has definite parity with respect to time-reversal and thus
\begin{multline}
\rho_{\mathcal{A}}(0\leq y_1<\dots<y_{n_A}\leq l;0\leq y'_1<\dots<y'_{n_A}\leq l)=
\\
\rho_{\mathcal{A}}(0\leq r_1<\dots<r_{n_A}\leq l;0\leq r'_1<\dots<r'_{n_A}\leq l)\delta_{\sigma_0,\sigma'_0}.
\end{multline}
It follows that the reduced density matrix in the sector with $n_A$ particles is a direct sum of 2 identical spin-sectors. This proves that the
entanglement spectrum must be doubly degenerate.

\subsection{Luttinger correlation functions}
Given that the wave-function of our system can be related to the wave-function of spin-less fermions on a lattice with attractive interactions
(Eq.~\eqref{eq:psi}), we expect to be able to relate correlation functions
to the spin-less fermion correlation functions which are known from
bosonization of the Luttinger model. The correlation functions
of operators that preserve the antiferromagnetic spin structure (i.e
do not introduce spin defects) can be mapped to correlation
functions of the spin-less Luttinger model and therefore turn
out to be gapless.
\subsubsection{Density-density correlation function}
The density correlation function is the simplest to connect to the
spin-less model. The density operator is defined by $\rho(r)=\sum_j \delta(r-r_j)$.The density-density correlator can be written
in terms of wave-functions as
\begin{multline}
\langle\rho(r)\rho(r')\rangle=\sum_{\mathbf y}|\Psi_0(\mathbf y)|\sum_{a,b}\delta(r-r_a)\delta(r'-r_b)
\\
=\sum_{\mathbf r}|\psi_0(\mathbf r)|\sum_{a,b}\delta(r-r_a)\delta(r'-r_b).
\end{multline}
The above expression is precisely the density-density correlation
function of spinless fermions on a lattice and is given by
\begin{equation}
\langle\rho(r)\rho(0)\rangle=\rho_0^2-\frac{K }{2\pi r^2}\frac{1-\frac{\lambda^2}{r^2}}{(1+\frac{\lambda^2}{r^2})^2}+\frac{2}{2\pi\lambda^2}\cos{(2 k_F r)}\left(\frac{\lambda}{r}\right)^{2 K}
\end{equation}
where $\rho_0 = \langle\rho(0)\rangle$ and $K>1$ is the Luttinger constant for spin-less fermions with
attractive interactions \cite{Giamarchi}. Here $\lambda$ is taken to be the
lattice scale.
\subsubsection{$S_z$ correlation functions}
The model discussed can also be thought of as a spin-1 Hamiltonian.
In this language, the natural correlation functions are the $S_{r,z}S_{r',z}$ correlation functions. The operator $S_{r,z}$ is defined as
\begin{equation}
S_{r,z}=\sum_j \sigma_j  \delta(r-r_j).
\end{equation}
The correlator is calculated  by evaluating
\begin{align}
\langle S_{r,z}S_{r',z}\rangle &=\sum_{\mathbf y}|\Psi(\mathbf y)|\sum_{a,b}\sigma_a\sigma_b\delta(r-r_a)\delta(r'-r_b)\nonumber\\
&=\sum_{\mathbf r}|\psi(\mathbf r)|\sum_{a\leq b}e^{i\pi(a-b)}\delta(r-r_a)\delta(r'-r_b)\nonumber\\
&=\langle \rho(r)e^{i\pi\sum_{r<p<r'}\rho(p)}\rho(r')\rangle
\end{align}
where the last correlator is for the spin-less fermions.
Here for the second to last statement we have used the fact that the spin
configuration is antiferromagnetic and therefore the spins are opposite
if there are an even number of particles between a and b.
Following Ref.~\cite{Giamarchi}, we use
\begin{equation}\label{eq:rho}
\rho(x) = \rho_0 - \frac{1}{\pi} \partial_x\phi(x) + \frac{1}{2\pi\lambda} \left[e^{-2 i\left(k_F x - \phi\left(x\right)\right)} + e^{2 i\left(k_F x - \phi\left(x\right)\right)} \right]
\end{equation}
in the string to get
\begin{equation}
\langle S_{r,z}S_{r',z}\rangle=\langle \rho(r) e^{-i\left( \phi\left(r'\right)-\phi\left(r\right) -k_F\left(r'-r\right) \right)}\rho(r')\rangle
\end{equation}
where for the small lattice constant limit $k_F a\ll 1$ we have ignored
the back-scattering term. Here we also have to be careful to
take only the real part of this expression.
Substituting Eq.~\eqref{eq:rho} into the above to express everything in terms of $\phi$ leads to the expression
\begin{widetext}
\begin{multline}
\langle S_{r,z}S_{0,z}\rangle = \rho_0^2 e^{-i k_F r} \langle e^{i\left(\phi(r)-\phi(0)\right)} \rangle + \left(\rho_0+ \frac{1}{2\pi\lambda}\right) \frac{e^{i k_F r}}{2\pi\lambda} \langle e^{-i\left(\phi(r)-\phi(0)\right)} \rangle
+ \left(\rho_0+ \frac{1}{2\pi\lambda}\right) \frac{e^{-i 3k_F r}}{2\pi\lambda} \langle e^{i3\left(\phi(r)-\phi(0)\right)}\rangle
\\
+ \frac{e^{-i k_F r}}{\pi^2}\langle \partial_r \phi(r) e^{i\left(\phi(r)-\phi(0)\right)} \partial_r \phi(0)\rangle
- \frac{\rho_0 e^{-i k_F r}}{\pi} \langle \left(\partial_r \phi(r)+\partial_r \phi(0)\right) e^{i\left(\phi(r)-\phi(0)\right)} \rangle
\\
=
\rho_0^2 e^{-i k_F r} \langle e^{i\left(\phi(r)-\phi(0)\right)} \rangle + \left(\rho_0+ \frac{1}{2\pi\lambda}\right) \frac{e^{i k_F r}}{2\pi\lambda} \langle e^{-i\left(\phi(r)-\phi(0)\right)} \rangle
+ \left(\rho_0+ \frac{1}{2\pi\lambda}\right) \frac{e^{-i 3k_F r}}{2\pi\lambda} \langle e^{i3\left(\phi(r)-\phi(0)\right)} \rangle
\\
+ \frac{e^{-i k_F r}}{\pi^2}\partial_r^2\langle e^{i\left(\phi(r)-\phi(0)\right)} \rangle
- \frac{2 \rho_0 e^{-i k_F r}}{\pi} \partial_r \langle e^{i\left(\phi(r)-\phi(0)\right)} \rangle.
\end{multline}
\end{widetext}
The slowest decay occurs from the first two terms and we get
\begin{equation}
\langle S_{r,z}S_{0,z}\rangle\sim \left[\rho_0^2+\frac{\rho_0}{2\pi\lambda}+\frac{1}{\left(2\pi\lambda\right)^2}\right] \cos{(k_F r)}\left(\frac{\lambda}{r}\right)^{K/2}.
\end{equation}
\subsubsection{$S_x$ string correlator}
In a previous section, we found that the ground state was
string ordered in the $S_z$ string correlations. A natural question
to ask is whether this system is also string ordered along $S_x$
similar to the Haldane model. To assess this we study the correlation
function of the string operator along $x$
\begin{equation}
O_{\text{str}}^{x}(r<r')=S_{r,x}\prod_{r<p<r'}e^{i\pi S_{p,x}}S_{r',x}.
\end{equation}
The spin-1 matrix $S_x$ is not diagonal in spin space. Instead it
converts filled sites to empty sites in the fermion representation and
empty sites are converted into the superposition $|+\rangle+|-\rangle$.
Since $S_x$ changes the number of particles, the action of $O_{\text{str}}^{x}$ on
$\Psi_0$ is non-zero only if exactly one of the sites $r$ and $r'$
is filled.
The factor $e^{i\pi S_{x}}$ apart from an overall $-$ sign,
flips all spins and leaves empty sites unchanged. Therefore the
action of $O_{\text{str}}^{x}$ on an antiferromagnetic ground state $|\Psi\rangle$
adds a particle at $r$ or $r'$ and removes a particle from the other
site, while flipping the spins in the middle to retain the
antiferromagnetic order.
 Therefore in the spin-less fermion language the string correlator
can be written as
\begin{equation}
\langle O_{\text{str}}^{x}(r<r')\rangle=(-1)^{(r-r')}\langle c^\dagger(r) e^{i\pi\sum_{r<p<r'}\rho(p)}c(r')\rangle.
\end{equation}
Here one must keep the real part only.
As before ignoring back scattering terms in the exponent
\begin{equation}
\langle O_{\text{str}}^{x}(r<r')\rangle=\langle c^\dagger(r) e^{-i\left( \phi\left(r'\right)-\phi\left(r\right) -k_F\left(r'-r\right) \right)} c(r')\rangle.
\end{equation}
From Ref.~\cite{Giamarchi}, $c(r)=e^{-i\theta(r)}\cos{\left(\phi(r)-k_F r\right)}$. Using this expression,
\begin{multline}
\langle O_{\text{str}}^{x}(0<r)\rangle= \frac{(-1)^r}{4} \sum_{s=\pm 1} \langle e^{-i\left(s \phi(0)-\phi(0)+\theta(r)-\theta(0)\right)}\rangle
\\
+ \frac{(-1)^r}{4} e^{i 2k_F r}\sum_{s=\pm 1} \langle e^{-i\left(2\phi(r) + s \phi(0)-\phi(0)+\theta(r)-\theta(0)\right)}\rangle.
\end{multline}
Evaluating the expectation values, the leading power-law behavior comes from the first term with $s=+1$:
\begin{equation}
\langle O_{\text{str}}^{x}(0<r)\rangle=\frac{(-1)^r}{4} \left(\frac{\lambda}{r}\right)^{1/2 K}.
\end{equation}
\subsection{Gapped correlation functions}
From our discussion of $O_{\text{str}}^{x}$, the product $S_{r,x}S_{r',x}$ clearly
is non-zero only when it acts on configurations where exactly one of $r$ and $r'$ is empty. Furthermore,
applying $S_{r,x}S_{r',x}$ to the antiferromagnetically aligned state $\Psi$ introduces a defect at the filled site (since it empties the site) unless all sites between $r$ and $r'$ are
empty. The probability of this decays exponentially with an exponent proportional to $\rho_0 |r-r'|$.  The fermion correlation function in the spinful fermion model is similarly localized.


\begin{thebibliography}{99}
\bibitem{note} The analytic expression for the phase boundary can be obtained from a mean-field treatment (after making a nonlocal unitary transformation as in Ref.~\citep{Kennedy92}).
\end{thebibliography}
\end{document}